\documentclass[%
reprint,
superscriptaddress,
 amsmath,amssymb,
 aps,
prb,
]{revtex4-2}

\usepackage{graphicx}
\usepackage{bm}
\usepackage[draft]{changes}
\usepackage{hyperref}
\hypersetup{
    colorlinks=true,
    linkcolor=blue,
    citecolor=blue,     
    urlcolor=blue,
}

\allowdisplaybreaks

\begin{document}

\title{Tracking unconventional superconductivity in the presence of strongly correlated Fermi arcs}



\author{Xianliang Zhou}
 \affiliation{%
 Beijing National Laboratory for Condensed Matter Physics and Institute of Physics, Chinese Academy of Sciences, Beijing 100190, China
}%
\author{Fei Yang}
\affiliation{Department of Materials Science and Engineering and Materials Research Institute, The Pennsylvania State University, University Park, PA 16802, USA}
\author{Miao Liu}
 \affiliation{%
 Beijing National Laboratory for Condensed Matter Physics and Institute of Physics, Chinese Academy of Sciences, Beijing 100190, China
}%
\author{Yin Shi}
 \email{yin.shi@iphy.ac.cn}
 \affiliation{%
 Beijing National Laboratory for Condensed Matter Physics and Institute of Physics, Chinese Academy of Sciences, Beijing 100190, China
}%
\author{Sheng Meng}
 \email{smeng@iphy.ac.cn}
 \affiliation{%
 Beijing National Laboratory for Condensed Matter Physics and Institute of Physics, Chinese Academy of Sciences, Beijing 100190, China
}%

\date{\today}

\begin{abstract}
One of the primary reasons that superconductivity in underdoped cuprates is enigmatic is that it emerges from an incoherent Fermi-arc state, so the applicability of the Bardeen–Cooper–Schrieffer (BCS) theory is questionable.
Here we approach this problem by investigating unconventional $d$-wave superconductivity in a recently proposed solvable model for strongly correlated Fermi arcs.
We show analytically that the exact incorporation of Fermi arcs fundamentally modifies the BCS equations, which enables us to isolate a many-body effect that suppresses the superconducting transition temperature $T_c$ beyond the simple reduction expected from a shrinking Fermi surface. 
The theory unambiguously produces: (i) a $T_c$ tracing out a dome as a function of hole doping, (ii) a new low-energy mode upon entering superconductivity, (iii) a suppressed superfluid stiffness in the underdoped regime, and (iv) a gap-to-$T_c$ ratio far exceeding the BCS limit, all consistent with experimental observations in cuprate superconductors.
These findings provide an analytic benchmark for understanding how superconductivity emerges from a correlated Fermi-arc state in high-$T_c$ superconductors.
\end{abstract}
\maketitle

\section{Introduction}In underdoped cuprate superconductors, the normal state is a correlated metal whose Fermi surface is not a single closed contour but is partly incoherent, leaving only separate segments known as Fermi arcs~\cite{lee06doping,keimer_quantum_2015,huscroft01pseudogaps, maier02angle-resolved, maier03fermi, parcollet04cluster, civelli05dynamical, macridin06pseudogap, kyung06pseudogap, stanescu06fermi, haule07strongly, civelli08nodal-antinodal, gull08local, werner09momentumselective, gull09momentum-sector-selective, gull10momentum-space, kuchinskii12generalized, gull12energetics, sordi12strong, gull13superconductivity, leblanc13equation, gunnarsson15fluctuation, wu18pseudogap, schäfer21tracking, ye23location, šimkovic24origin}.
The presence of Fermi arcs reduces the density of states at the Fermi level, leading to the emergence of a pseudogap.
A central open question is how superconductivity is built in the presence of the strongly correlated Fermi arcs and how Fermi-arc (pseudogap) physics shapes the superconducting properties. 
The conventional Bardeen--Cooper--Shrieffer (BCS) theory essentially works only for superconductivity emerging from coherent Fermi liquids, so it is not enough to resolve this question.

Arguably part of the Fermi-arc phenomenon may stem from preformed Cooper pairs that have not yet developed global phase coherence~\cite{norman07modeling,reber12the}, suggesting a cooperative relationship between the arcs and superconductivity. 
Nonetheless, it is generally believed that additional mechanisms are involved.
Overall, the main influence of Fermi arcs on superconductivity is expected to be detrimental, as the effective Fermi surface is reduced so that the density of electrons that can participate in pairing is lowered~\cite{norman05the,mishra14an}—an effect that can be described as a spectral suppression of the superconducting transition temperature $T_c$. 
This is consistent with the observed negative correlation between $T_c$ and the pseudogap onset temperature $T^*$~\cite{keimer_quantum_2015}. 
On the other hand, numerical investigations of the hole-doped Hubbard model found that the pairing interaction between antinodal fermions mediated by spin fluctuations is substantially enhanced when the pseudogap is present, relative to the full metallic phase~\cite{Maier16pairing,yang25pairing}. 
This enhancement helps superconductivity persist within the pseudogap regime, where it might otherwise be absent~\cite{mishra14an}.

A natural question is then whether Fermi arcs have additional consequences for superconductivity.
This is directly tied to how $T_c$ in materials can be optimized.
If such effects do exist, they are challenging to be disentangled, since the most basic theoretical description of cuprates—the square-lattice Hubbard model—is intractable.
It is thus advantageous to investigate a tractable model that features both Fermi arcs and superconductivity.

In this work, we investigate $d$-wave superconductivity in a recently proposed solvable Hamiltonian that generates Fermi arcs through a transparent microscopic mechanism~\cite{worm24fermi}.
By supplementing this model with a $d$-wave pairing interaction, plausibly originating from the antiferromagnetic exchange interaction, we derive an asymptotically exact pairing susceptibility and obtain $T_c$ over the full doping range.
The resulting phase diagram exhibits a characteristic superconducting dome, with optimal doping located near the quantum critical point separating the pseudogap regime from the fully metallic phase.
The solvability of the model further enables us to identify analytically a many-body effect associated with the Fermi arcs that suppresses $T_c$ beyond the simple reduction expected from the loss of Fermi-level density of states alone.
We further show that this many-body effect gives rise to a new low-energy mode upon entering superconductivity, a pronounced suppression of the superfluid stiffness, and a gap-to-$T_c$ ratio far exceeding the BCS limit in the underdoped regime, all of which are consistent with experimental observations in cuprate superconductors.
These results provide an analytic benchmark for understanding how superconductivity emerges from a correlated Fermi-arc state in high-$T_c$ superconductors.

\section{The Fermi-arc model}We begin by introducing the solvable model that captures the Fermi-arc physics underlying our analysis. The Hamiltonian is~\cite{worm24fermi}:
\begin{equation}
\hat{H}_0=\sum_{\mathbf{k} \sigma}\left(\xi_{\mathbf{k}} \hat{n}_{\mathbf{k} \sigma}+\frac{U}{2} \hat{n}_{\mathbf{k} \sigma} \hat{n}_{\mathbf{k}+\mathbf{Q}\bar{\sigma}}\right),
\label{eqH}
\end{equation}
where $\hat{n}_{\mathbf{k} \sigma}=\hat{c}_{\mathbf{k} \sigma}^{\dagger} \hat{c}_{\mathbf{k} \sigma}$ is the number operator with $\hat{c}_{\mathbf{k} \sigma}$ ($\hat{c}_{\mathbf{k} \sigma}^{\dagger}$) being the annihilation (creation) operators for an electron with momentum $\mathbf{k}$ and spin $\sigma$. 
Here $U > 0$ is a repulsive interaction, and $\mathbf{Q}=(\pi, \pi)$ is the characteristic wave vector of antiferromagnetic fluctuations~\cite{worm24fermi}. 
The single-particle dispersion is
\begin{equation}
\begin{split}
\xi_\mathbf{k}=&  -2 t\left[\cos \left(k_x\right)+\cos \left(k_y\right)\right]-4 t^{\prime} \cos \left(k_x\right) \cos \left(k_y\right)\\
 &-2 t^{\prime \prime}\left[\cos \left(2 k_x\right)+\cos \left(2 k_y\right)\right]-\mu,
 \end{split}
\end{equation}
where $\mu$ is the chemical potential and $t \equiv 1$ (setting the energy unit, which is around $0.3$ eV in cuprates), $t^{\prime}=-0.2$, and $t^{\prime \prime}=0.1$ are set to typical values used in low-energy tight-binding models for cuprates \cite{nicoletti10high, worm2023numerical}.

Compared with the Hubbard interaction $(U/2) \sum_{\mathbf{k k' q \sigma}} \hat{c}^\dagger_{\mathbf{k}\sigma} \hat{c}_{\mathbf{k}'\sigma} \hat{c}^\dagger_{\mathbf{k}'+\mathbf{q}\bar{\sigma}} \hat{c}_{\mathbf{k}+\mathbf{q}\bar{\sigma}}$, the interaction term in Eq.~(\ref{eqH}) is restricted to the dominant coupling corresponding to the long-range antiferromagnetic spin fluctuations.
$H_0$ is nearly diagonal in momentum space (as it only couples $\mathbf{k}$ and $\mathbf{k}+\mathbf{Q}$ states) and its retarded Green's function is readily available:
\begin{equation}
G_{0\sigma}(\mathbf{k},\omega)=\frac{1-n_{\mathbf{k}+\mathbf{Q}}}{\omega-\xi_\mathbf{k}+\mathrm{i} 0^{+}}+\frac{n_{\mathbf{k}+\mathbf{Q}}}{\omega-\xi_\mathbf{k}-U+\mathrm{i} 0^{+}},
\label{eqG}
\end{equation}
where $n_{\mathbf{k}+\mathbf{Q}}=\left\langle\hat{n}_{\mathbf{k}+\mathbf{Q} \sigma}\right\rangle$ is the electron occupation (assuming no magnetic order) given by:
\begin{equation}
n_{\mathbf{k}+\mathbf{Q}}=\frac{e^{-\beta \xi_{\mathbf{k}+\mathbf{Q}}}+e^{-\beta\left(\xi_\mathbf{k}+\xi_{\mathbf{k}+\mathbf{Q}}+U\right)}}{1+e^{-\beta \xi_\mathbf{k}}+e^{-\beta \xi_{\mathbf{k}+\mathbf{Q}}}+e^{-\beta\left(\xi_\mathbf{k}+\xi_{\mathbf{k}+\mathbf{Q}}+U\right)}},
\label{eqnkq}
\end{equation}
where $\beta=1/T$, with $T$ being the temperature in units of energy.
The corresponding spectral function is
\begin{equation}
A(\mathbf{k}, \omega)=\left(1-n_{\mathbf{k}+\mathbf{Q}}\right) \delta\left(\omega-\xi_\mathbf{k}\right)+n_{\mathbf{k}+\mathbf{Q}} \delta\left(\omega-\xi_\mathbf{k}-U\right).
\label{eqspectrum}
\end{equation}

\begin{figure}
\includegraphics[width=8.6cm]{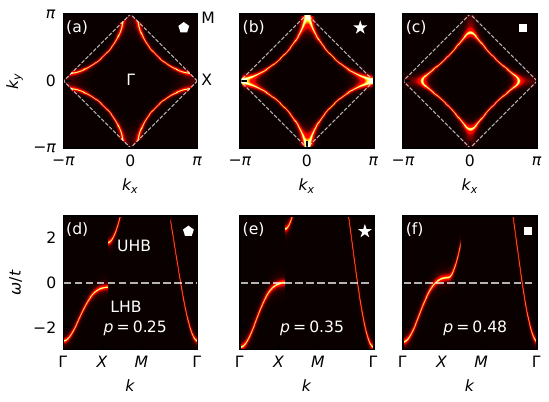}
\centering
\caption{\label{fig1}(a) Normal-state Fermi surface and (d) single-particle excitation spectrum at a hole doping of $p=0.25$ for $U=2$ at $T= 0$. The dashed lines in (a) are the antiferromagnetic zone boundary. 
UHB and LHB are short for upper and lower Hubbard bands, respectively. Panels (b), (c), (e), and (f) are the same as (a) and (d) but for heavier doping levels of $p=0.35$ and $p=0.48$. An energy smearing of $0.04$ is used for visualization. The markers in the upper right corner denote the representative doping levels in Fig.~\ref{fig2}(a) and Fig.~\ref{fig3}(a).}
\label{figspectrum}
\end{figure}

Figure~\ref{figspectrum} shows the normal-state Fermi surfaces and corresponding spectral functions for three representative doping levels at $U=2$ and $T=0$ [these states are also marked in Fig.~\ref{fig2}(a)].
The hole concentration is defined by $p=1-(2/N)\sum_\mathbf{k}n_\mathbf{k}$ where $N$ is the number of lattice sites.
For the underdoped state ($p=0.25$), the noninteracting Fermi surface intersects with the antiferromagnetic zone boundary (AFZB), and the momentum-selective spectral weight defined through $n_{\mathbf{k}+\mathbf{Q}}$ transforms the Fermi surface outside the AFZB into a Luttinger surface [Fig.~\ref{figspectrum}(a)]~\cite{worm24fermi,Dzyaloshinskiiluttinger} and shifts the spectrum in these regions to energies above the Fermi level [Fig.~\ref{figspectrum}(d)].
The result is a disconnected Fermi surface, namely Fermi arcs, whose morphology [Fig.~\ref{figspectrum}(a)] resembles the experimentally observed normal-state Fermi surfaces in underdoped cuprate superconductors~\cite{damascelli2003angle,Vishik_2018,lee_abrupt_2007,hashimoto_energy_2014,vishik_phase_2012,he_single-band_2011}.
Note that the origin of Fermi arcs in this model is distinct from the mechanism involving preformed, incoherent Cooper pairs.

As the hole concentration is raised to $p=0.35$, the Fermi surface contracts so that it lies completely within the AFZB and thus remains intact [Fig.~\ref{figspectrum}(b)].
The Fermi level is now touching the lower Hubbard band edge at the antinodal points, thus forming the van Hove singularity [Fig.~\ref{figspectrum}(e)].
This identifies the quantum critical point separating the pseudogap phase from the fully metallic phase. For the overdoped state ($p=0.48$), the pseudogap disappears at the Fermi level [Fig.~\ref{figspectrum}(c) and (f)].
The Fermi arcs elongate with rising temperature, and the pseudogap disappearance (or onset) temperature $T^*$ is determined by the closing of the Fermi surface~\footnote{To determine whether the Fermi surface is closed, we employ an energy broadening of $\delta = 0.005$ and adopt a threshold for the spectral weight of $0.1/(\pi \delta)$, corresponding to $10\%$ of the maximum spectral intensity. Only spectral-weight values exceeding this threshold are regarded as non-empty.}.
The calculated $T^*$ decreases approximately linearly with increasing hole concentration [Fig.~\ref{fig2}(a)], consistent with the experimentally measured dependence $T^*\propto-p$ in cuprates~\cite{timusk_pseudogap_1999,armitage_progress_2010,keimer_quantum_2015,damascelli2003angle}.

\section{Superconducting transition temperature}
To analytically study superconductivity in the presence of Fermi arcs, we append the Hamiltonian in Eq. (\ref{eqH}) with an additional $d$-wave pairing interaction that could originate from the antiferromagnetic exchange interaction:
\begin{equation}
\hat{H}=\hat{H}_0-\frac{J}{N}\hat{\Delta}^{\dagger} \hat{\Delta},
\label{eqsuperconducting}
\end{equation}
where $J > 0$ is the exchange interaction strength. 
$\hat{\Delta}^{\dagger}=\sum_\mathbf{k} g_\mathbf{k} \hat{b}^{\dagger}_\mathbf{k}=\sum_\mathbf{k}g_\mathbf{k}\hat{c}^{\dagger}_{-\mathbf{k} \downarrow} \hat{c}^{\dagger}_{\mathbf{k} \uparrow}$, where $g_{\mathbf{k}}=\cos k_x-\cos k_y$ is the form factor of the $d$-wave pairing channel.
Equation~(\ref{eqsuperconducting}) can be regarded as a simplified $t$--$J$--$U$ model, in which the onsite Coulomb repulsion is restricted to certain momentum scattering channels [Eq.~(\ref{eqH})] and the exchange interaction is symmetrized to the $d$-wave channel.

The solvability of the Fermi-arc model permits an asymptotically exact formulation
of the pair susceptibility that accounts for the Fermi arcs exactly.
The pair susceptibility is defined as
\begin{align}
\chi(\tau)&=\frac{\operatorname{Tr}[\mathrm{e}^{-\beta \hat{H}} \mathcal{T} \hat{\Delta}_H(\tau) \hat{\Delta}^{\dagger}]}{N \operatorname{Tr}[\mathrm{e}^{-\beta \hat{H}}]}, \\
&= \frac{\operatorname{Tr}[\mathrm{e}^{-\beta \hat{H}_0}\mathcal{T} \hat{S}(\beta,0) \hat{\Delta}(\tau) \hat{\Delta}^\dagger]}{N \operatorname{Tr}[\mathrm{e}^{-\beta \hat{H}_0} \hat{S}(\beta, 0)]}, \label{eq:chi_def}
\end{align}
where the subscript $H$ of a time-dependent operator indicates the Heisenberg picture evolving under $\hat{H}$, while its absence defaults to the interaction picture evolving under $\hat{H}_0$.
$\mathcal{T}$ is the time ordering operator and $\hat{S}(\beta,0) = \sum_{m=0}^\infty \frac{J^m}{N^m m!} \mathcal{T} [ \int_0^\beta \mathrm{d}\tau \hat{\Delta}^\dagger(\tau) \hat{\Delta}(\tau) ]^m$ arises from the change of picture.
$\hat{H}_0$ is nearly diagonal in momentum space, $\hat{H}_0 = \sum_{\mathbf{k}\in\mathrm{HBZ}}\sum_\sigma \hat{H}_{\mathbf{k} \sigma}$ (HBZ denotes half Brillouin zone), where $\hat{H}_{\mathbf{k}\sigma} = \xi_{\mathbf{k}} \hat{n}_{\mathbf{k}\sigma} + \xi_{\mathbf{k}+\mathbf{Q}} \hat{n}_{\mathbf{k}+\mathbf{Q}\bar{\sigma}} + U\hat{n}_{\mathbf{k}\sigma} \hat{n}_{\mathbf{k}+\mathbf{Q}\bar{\sigma}}$. 
Notice that $\hat{H}_{\mathbf{k}' \sigma}$ commutes with the time-dependent pair operator $\hat{b}_\mathbf{k}(\tau) = \mathrm{e}^{-2[\xi_{\mathbf{k}} + \xi_{\mathbf{k}+\mathbf{Q}} + U(\hat{n}_{-\mathbf{k}+\mathbf{Q}\uparrow} + \hat{n}_{\mathbf{k}+\mathbf{Q}\downarrow})]\tau}\hat{b}_{\mathbf{k}}$, except for coincident cases $\mathbf{k}'=\mathbf{k}, \mathbf{k}+\mathbf{Q}$. 
Plugging $\hat{S}(\beta,0)$, $\hat{H}_0$, and $\hat{\Delta}(\tau)$ into Eq.~(\ref{eq:chi_def}) and using this commutativity (the coincident terms are subextensive and vanish in the thermodynamic limit), one sees that $\chi(\tau)$ can be simplified to a convolution of the bare susceptibility $\chi_0(\tau)$~\cite{phillips2020exact},
\begin{equation}
    \begin{aligned}
        \chi(\tau) ={}& \sum_{m=0}^\infty J^m \int_0^\beta \mathrm{d}\tau_1 \cdots \int_0^\beta \mathrm{d}\tau_m \chi_0(\tau-\tau_1) \\
        &\times \chi_0(\tau_1-\tau_2) \cdots \chi_0(\tau_{m-1} - \tau_m) \chi_0(\tau_m),
    \end{aligned} \label{eq:chitau}
\end{equation}
where
\begin{equation}
    \chi_0(\tau) = \sum_{\mathbf{k}} \frac{\gamma_{\mathbf{k}}^2 \operatorname{Tr}[\mathrm{e}^{-\beta (\hat{H}_{\mathbf{k}\uparrow}+\hat{H}_{-\mathbf{k}\downarrow})} \mathcal{T} \hat{b}_{\mathbf{k}}(\tau)\hat{b}_{\mathbf{k}}^\dagger]}{N \operatorname{Tr}[\mathrm{e}^{-\beta (\hat{H}_{\mathbf{k}\uparrow}+\hat{H}_{-\mathbf{k}\downarrow})}]}.
\end{equation}
The bare susceptibility can be further transformed to
\begin{equation}
    \chi_0(\tau) = \frac{1}{N} \sum_{\mathbf{k}} g_{\mathbf{k}}^2 G_{0\downarrow}(-\mathbf{k},\tau) G_{0\uparrow}(\mathbf{k},\tau), \label{eq:chi0tau}
\end{equation}
up to a thermodynamically vanishing contribution from coincident cases $-\mathbf{k}=\mathbf{k}, \mathbf{k}+\mathbf{Q}$.

Fourier transforming Eqs.~(\ref{eq:chitau},~\ref{eq:chi0tau}) to the Matsubara frequency ($\nu_n=2n\pi/\beta$) domain gives
\begin{align}
    \chi(\mathrm{i}\nu_n) &= \sum_{m=0}^\infty J^m \chi_0^{m+1}(\mathrm{i}\nu_n) = \frac{\chi_0(\mathrm{i}\nu_n)}{1-J\chi_0(\mathrm{i}\nu_n)}, \label{eq:chinu} \\
    \chi_0(\mathrm{i}\nu_n=0) &= \frac{1}{N}\sum_{\mathbf{k}}g_{\mathbf{k}}^2\int_{-\infty}^\infty \mathrm{d}\omega \frac{\tilde{A}(\mathbf{k},\omega)}{2\omega} \tanh\left(\frac{\beta\omega}{2}\right), \label{eq:chi0nu}
\end{align}
with $T_c$ determined by the divergence of $\chi(\mathrm{i}\nu_n=0)$,
\begin{equation}
    \chi_0(\mathrm{i}\nu_n=0) = \frac{1}{J}. \label{eqTc}
\end{equation}
Here the effective spectral function has two parts $\tilde{A}(\mathbf{k},\omega)=A(\mathbf{k},\omega)+A'(\mathbf{k},\omega)$, where 
\begin{equation}
A' = \frac{Z_{1\mathbf{k}}Z_{2\mathbf{k}}(\xi_{2\mathbf{k}}-\xi_{1\mathbf{k}})}{\xi_{1\mathbf{k}}+\xi_{2\mathbf{k}}} [\delta(\omega-\xi_{2\mathbf{k}}) - \delta(\omega-\xi_{1\mathbf{k}})], \label{eq:A'}
\end{equation}
is a many-body correction to the standard single-particle excitation spectrum $A$ for determining $T_c$.
$Z_{1\mathbf{k}}=1-n_{\mathbf{k}+\mathbf{Q}}$ and $\xi_{1\mathbf{k}}=\xi_\mathbf{k}$ are the spectral weight and energy of the lower Hubbard band, respectively, and $Z_{2\mathbf{k}}=n_{\mathbf{k}+\mathbf{Q}}$ and $\xi_{2\mathbf{k}}=\xi_\mathbf{k} + U$ are those of the upper Hubbard band.

We first conduct a qualitative analysis of Eq.~(\ref{eq:A'}).
If $A'$ is dropped from Eq.~(\ref{eq:chi0nu}), Eq.~(\ref{eqTc}) would be the standard BCS mean-field equation for determining the $d$-wave superconducting $T_c$, where only the single-particle spectral effect of the Fermi arcs on superconductivity is included.
First, for $\omega$ around the Fermi level, $A'(\mathbf{k},\omega)$ is nonzero only for $\mathbf{k}$ around the antinodal points, namely where the pseudogap is located.
This is because for $\mathbf{k}$ near the Fermi surface, $0<n_{\mathbf{k}+\mathbf{Q}}<1$ occurs only near the antinodal points, such that $Z_{1\mathbf{k}}Z_{2\mathbf{k}}$ [Fig.~\ref{figspectrum}(d)], and thus $A'(\mathbf{k},\omega)$, are nonzero only there.
Second, unlike the spectral function $A(\mathbf{k},\omega)$ that is non-negative, $A'(\mathbf{k},\omega)$ can be either positive or negative.
Since $\xi_{2\mathbf{k}}-\xi_{1\mathbf{k}}$ is always positive, the sign of $A'(\mathbf{k},\omega)$ is determined by $\xi_{1\mathbf{k}}+\xi_{2\mathbf{k}}$.
When the Fermi level lies below (above) the midpoint of the pseudogap at the antinodal points, one has $\xi_{1\mathbf{k}}+\xi_{2\mathbf{k}} > 0$ ($\xi_{1\mathbf{k}}+\xi_{2\mathbf{k}} < 0$) there, and meanwhile $\xi_{1\mathbf{k}}$ ($\xi_{2\mathbf{k}}$) governs the electronic behavior more strongly than $\xi_{2\mathbf{k}}$ ($\xi_{1\mathbf{k}}$).
This means that $A'(\mathbf{k},\omega)$ always adjusts $A(\mathbf{k},\omega)$ downward near the Fermi level, which will result in an additional decrease of $T_c$.
This effect can be termed many-body suppression of superconductivity originating from Fermi arcs.

\begin{figure}
\includegraphics[width=8.6cm]{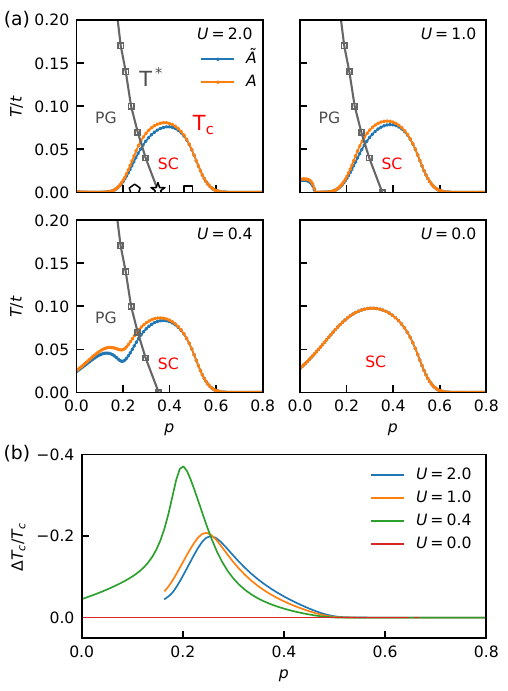}
\centering
\caption{\label{fig2}(a) Phase diagrams for several repulsion strengths at $J=0.8$. The gray line indicates the pseudogap (PG) onset temperature $T^*$, and the blue (orange) line represents the superconducting (SC) transition temperature $T_c$ calculated with (without) the many-body correction $A^{\prime}$. The momentum summation is carried out on a $512 \times 512$ $\mathbf{k}$-grid. The markers correspond to underdoped, optimally doped and overdoped states, whose normal-state Fermi surfaces are shown in Fig.~\ref{figspectrum}. (b) Relative difference between the superconducting transition temperatures calculated with and without $A^{\prime}$.}
\end{figure}

We obtain $T_c$ by numerically solving Eq.~(\ref{eqTc}). 
Figure~\ref{fig2}(a) shows the phase diagrams for several repulsions $U$ at a fixed pairing potential $J=0.8$ with and without the many-body correction $A'$. 
For $U=2$, the normal state at half filling ($p=0$) is a fully gapped insulator, and $T_c$ exhibits a characteristic dome-shaped dependence on doping concentration, both with and without $A'$.
This is primarily a consequence of the dependence of $A$ on doping concentration. 
When the system is doped to the quantum critical point separating the pseudogap and metallic regimes, the Fermi level aligns with the van Hove singularity, leading to the peak of the superconducting dome—the so-called optimal doping (star marker).
As the doping concentration is lowered, the Fermi arcs contract, progressively reducing the density of states at the Fermi level and thus suppressing $T_c$.
Once the Fermi arc length shrinks to a finite critical value, Eq.~(\ref{eqTc}) no longer admits a solution, resulting in $T_c = 0$ (pentagon marker)~\cite{mishra14an}. 
These features are compatible with the superconducting dome not extending to half filling in cuprate superconductors~\cite{timusk_pseudogap_1999,armitage_progress_2010,keimer_quantum_2015,damascelli2003angle}. 

For weaker values of the repulsion ($U=1$ or $0.4$), a smaller dome separate from the main superconducting dome appears as doping progresses to the electron-doped side because the normal state around half filling is now a hot-spot metal~\cite{worm24fermi}, which can appear in electron-doped cuprates~\cite{armitage02doping,matsui07evolution}.
The $U=0$ case just recovers the BCS theory.

Beyond the characteristic superconducting dome that arises primarily from the single-particle spectral properties of the Fermi arcs, Fig.~\ref{fig2} highlights the effect of the many-body correction $A^{\prime}$ on $T_c$. 
$A^{\prime}$ consistently suppresses $T_c$ across all repulsion strengths in the underdoped regime, consistent with our qualitative analysis above.  
Figure~\ref{fig2}(b) shows the relative difference between $T_c$ calculated with and without $A'$ for several interaction strengths, derived from the data in Fig.~\ref{fig2}(a). 
The most pronounced many-body suppression of superconductivity appears around the middle of the underdoped regime.
As $U$ increases, the maximal suppression ratio initially rises rapidly, but then declines and settles at a finite value.

\begin{figure}
\includegraphics[width=8.6cm]{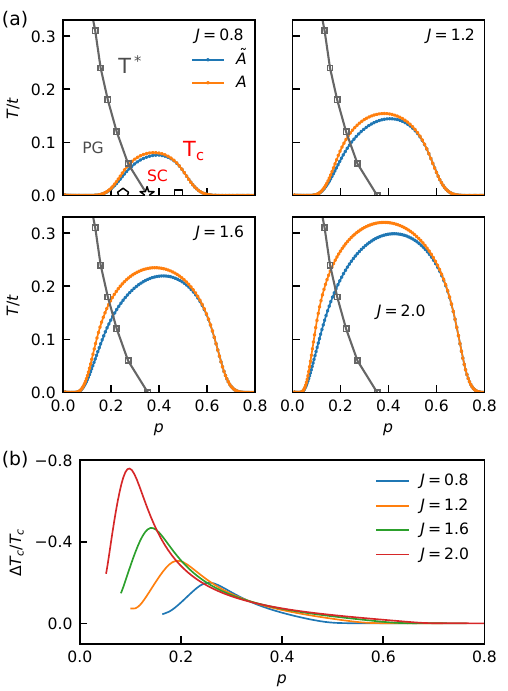}
\centering
\caption{\label{fig3}Same as Fig.~\ref{fig2} but for several pair interactions at a fixed repulsion strength $U=2$.}
\end{figure}

Figure~\ref{fig3} shows the phase diagram for several pairing interactions $J$ at a fixed repulsion $U=2$. 
As $J$ increases, the superconducting dome expands; however, on the low-doping side there remains a finite doping interval over which superconducting order is completely suppressed.
Moreover, increasing $J$ leads to a more pronounced many-body reduction of $T_c$ and moves the point of strongest suppression toward lower doping levels [Fig.~\ref{fig3}(b)].
This occurs because, as the temperature rises, the region in $\mathbf{k}$ space where $0 < n_{\mathbf{k}+\mathbf{Q}} < 1$ expands, so that $Z_{1\mathbf{k}}Z_{2\mathbf{k}}$, and consequently $A'(\mathbf{k},\omega)$, become nonzero for a larger set of $\mathbf{k}$ points.
This indicates that, as $T_c$ increases, many-body effects that suppress superconductivity play an increasingly important role in setting the value of $T_c$.
Nevertheless, the many-body suppression ratio of $T_c$ at optimal doping remains almost the same as $J$ varies.

In the doped Hubbard model and in cuprate materials, the single-particle spectrum in the vicinity of the antinodal points is significantly more incoherent than in the solvable model of Eq.~(\ref{eqH}), because there are many more available scattering channels~\cite{worm24fermi}.
Consequently, the region in $\mathbf{k}$ and $\omega$ space where $A'(\mathbf{k},\omega)$ is finite becomes much broader, implying that the many-body suppression of superconductivity arising from the Fermi arcs should be stronger than in the solvable model.

\begin{figure*}
\includegraphics[width=17.8cm]{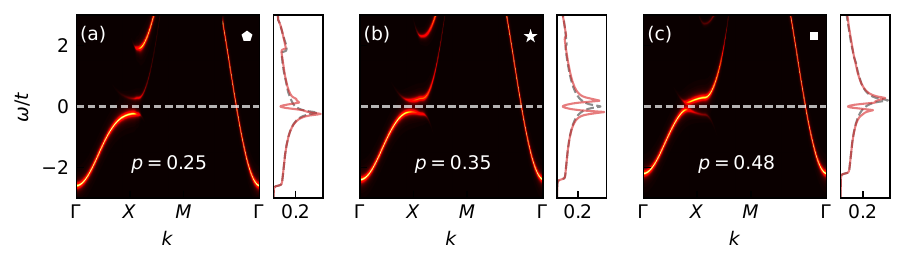}
\centering
\caption{\label{fig5}Single-particle spectral function and the corresponding density of states in the superconducting ground state for $U=2$ and $J=0.8$ at three representative doping levels, as in Fig.~\ref{fig1}. 
An energy smearing of $0.04$ is used for visualization. 
The dashed gray curves show the density of states in the corresponding normal state.}
\end{figure*}

\section{Superconducting spectral function}To characterize the single-particle excitations in the superconducting state, we calculate the corresponding spectral function by treating the pairing interaction $J$ at the mean-field level while retaining the repulsive interaction $U$ exactly. 
The resulting Hamiltonian is:
\begin{equation}
\begin{split}
\hat{H}_{\mathrm{MF}}={}&
\hat{H}_0
+\sum_{\mathbf{k}}
\left(
\Delta_{\mathbf{k}}\hat{b}_{\mathbf{k}}^{\dagger}
+\Delta_{\mathbf{k}}^{*}\hat{b}_{\mathbf{k}}
\right),
\label{HMF}
\end{split}
\end{equation}
where
\begin{equation}
\Delta_{\mathbf{k}}=-\frac{Jg_{\mathbf{k}}}{N} \sum_{\mathbf{k}'} g_{\mathbf{k}'}\langle \hat{b}_{\mathbf{k}'}\rangle.
\label{delta}
\end{equation}
By choosing a non-overlapping set of momenta, the Hamiltonian in Eq.~(\ref{HMF}) decomposes into independent momentum sectors. Each sector contains the eight modes:
\begin{align}
&(\mathbf{k}\uparrow,\mathbf{k}+\mathbf{Q}\downarrow,
-\mathbf{k}\downarrow,-\mathbf{k}-\mathbf{Q}\uparrow),\\
&(\mathbf{k}\downarrow,\mathbf{k}+\mathbf{Q}\uparrow,
-\mathbf{k}\uparrow,-\mathbf{k}-\mathbf{Q}\downarrow).
\end{align}
Owing to spin degeneracy, these modes further separate into two equivalent and mutually decoupled four-mode blocks. Each minimal block therefore has a Fock-space dimension of $2^4=16$ and can be solved by exact diagonalization. The order parameter $\Delta_{\mathbf{k}}$ is determined self-consistently from Eq.~(\ref{delta}). The single-particle spectral function is then obtained from the Lehmann representation:
\begin{equation}
A(\mathbf{k}, \omega)=\sum_{n m}|\langle n| \hat{c}_\mathbf{k}| m\rangle|^ 2\left(\rho_n+\rho_m\right) \delta\left(\omega+E_n-E_m\right),
\end{equation}
where $|n\rangle$, $|m\rangle$, $E_n$ and $E_m$ are eigenstates and eigenenergies of $\hat{H}_\text{MF}(\mathbf{k})$, $\rho_n= e^{-\beta E_n} / Z$ and $Z=\sum_n e^{-\beta E_n}$ is the partition function.

Figure~\ref{fig5} shows the single-particle spectral functions and the corresponding density of states in the superconducting ground state at three representative doping levels. 
In the underdoped regime [Fig.~\ref{fig5}(a)], where the normal state exhibits an antinodal pseudogap, the emergence of superconductivity not only backbends the lower Hubbard band near the antinodes, but also generates a new low-energy mode inside the pseudogap. 
This behavior is consistent with an ARPES measurement in underdoped cuprates showing that the emergence of superconductivity is associated with the formation of new low-lying states inside the preexisting pseudogap~\cite{he_single-band_2011}.
At the optimal doping point [Fig.~\ref{fig5}(b)], where the Fermi level coincides with the van Hove singularity, the emergence of superconductivity also produces an additional low-energy mode; however, this mode carries much more spectral weight than that in the underdoped regime.

In the overdoped regime [Fig.~\ref{fig5}(c)], where the normal-state spectrum is gapless and regular at the Fermi level, the emergence of superconductivity just opens a nodal gap in the regular band and produces two coherence peaks, consistent with the low-energy expectations of standard BCS theory.

\section{Superfluid stiffness}The pseudogap also affects the superfluid stiffness of the superconducting state. Within the mean-field treatment of the pairing interaction [Eq.~(\ref{HMF})], the superfluid stiffness $D_s$ is obtained from the static current response as~\cite{phillips2020exact}:
\begin{align}
\frac{D_s}{\pi} 
& =\frac{1}{N}\left(\left\langle \hat{K}_{x x}\right\rangle-\sum_{n m}|\langle n| \hat{J}_x| m\rangle|^ 2 \frac{\rho_m-\rho_n}{E_n-E_m}\right)
\label{eq:stiffness}
\end{align}
where
\begin{align}
\hat{K}_{xx}
&=\sum_{\mathbf{k}\sigma}
\frac{\partial^2 \xi_{\mathbf{k}}}{\partial k_x^2}
\hat{c}_{\mathbf{k}\sigma}^{\dagger}
\hat{c}_{\mathbf{k}\sigma}, \\
\hat{J}_x
&=\sum_{\mathbf{k}\sigma}
\frac{\partial \xi_{\mathbf{k}}}{\partial k_x}
\hat{c}_{\mathbf{k}\sigma}^{\dagger}
\hat{c}_{\mathbf{k}\sigma}.
\end{align}
The first and second terms in Eq.~(\ref{eq:stiffness}) are the diamagnetic and paramagnetic contributions, respectively.

\begin{figure}
\includegraphics[width=8.6cm]{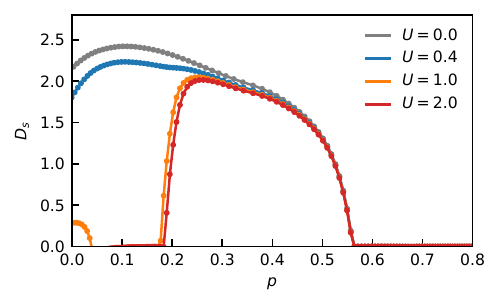}
\centering
\caption{\label{fig6}Superfluid stiffness as a function of doping for several repulsion strengths at $J=0.8$ and $T=0.01$. The momentum summation is carried out on a $1024 \times 1024$ $\mathbf{k}$-grid.}
\end{figure}

Figure~\ref{fig6} shows the superfluid stiffness for several repulsion strengths at fixed $J=0.8$. The superfluid stiffness is finite only within the superconducting region (Fig.~\ref{fig2}). Compared with the BCS limit ($U=0$), it is substantially suppressed in the pseudogap regime, reflecting the depletion of low-energy spectral weight available to carry the supercurrent as the Fermi surface is truncated into Fermi arcs. In the overdoped regime, although the normal-state spectrum remains gapless at the Fermi level, as in a conventional metal, the superfluid stiffness still decreases with increasing repulsion strength. This suppression cannot be attributed to a loss of Fermi-level spectral weight and therefore reveals a many-body effect of the repulsive interaction.

\section{Gap-to-$T_c$ ratio}To further characterize the unconventional superconductivity in the Fermi-arc regime, we examine the zero-temperature superconducting gap and the resulting gap-to-$T_c$ ratio. To calculate the superconducting gap at zero temperature, we use a BCS-type variational wavefunction $|\psi\rangle=\prod_{\mathbf{k}}[\cos(\theta_\mathbf{k})+\sin(\theta_\mathbf{k})\hat{b}_\mathbf{k}^{\dagger}]|0\rangle$, because the Fermi-arc model (\ref{eqH}) only has unoccupied and doubly occupied  $\mathbf{k}$-states at zero temperature.
Minimizing the energy $\langle\psi| \hat{H}|\psi\rangle$ with respect to $\theta_\mathbf{k}$ yields the following equation:
\begin{align}
&2\xi_{\mathbf{k}}\sin(2\theta_\mathbf{k})+2U\sin(2\theta_\mathbf{k})\sin^2(\theta_{\mathbf{k}+\mathbf{Q}}) \nonumber \\
&-\frac{J}{N}g_\mathbf{k}\cos(2\theta_\mathbf{k})\sum_{\mathbf{k'}}g_{\mathbf{k}'}\sin(2\theta_{\mathbf{k}'})=0. \label{eq:constrain}
\end{align}
With the zero-temperature (maximum, namely antinodal) superconducting gap defined by $\Delta(0)=(2J/N)\langle\psi|\hat\Delta|\psi\rangle=(J/N)\sum_{\mathbf{k}}g_{\mathbf{k}}\sin(2\theta_{\mathbf{k}})$, Eq.~(\ref{eq:constrain}) cannot be reduced to a single closed equation for $\Delta(0)$ due to the $U$ term.
We thus need to explicitly solve the set of equations for $\theta_\mathbf{k}$.
If $U=0$, Eq.~(\ref{eq:constrain}) falls back to the BCS gap equation and can be reduced to one closed equation for $\Delta(0)$.

\begin{figure}
\includegraphics[width=8.6cm]{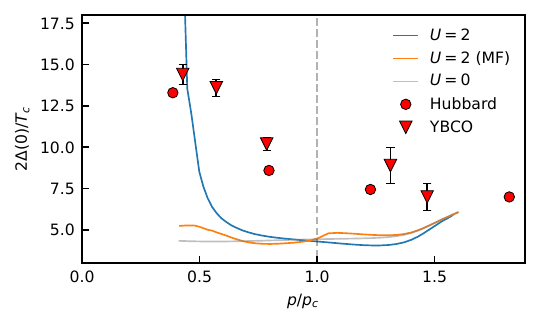}
\centering
\caption{\label{fig4} Ratio of the zero-temperature superconducting gap to the transition temperature, $2\Delta(0)/T_c$, as a function of doping. The curves are the results from this model (MF: mean field) for $J=0.8$. The red circles and triangles are numerical results for the Hubbard model~\cite{gull13superconductivity} and experimental data for YBCO superconductors~\cite{inosov11crossover}, respectively. The doping level is normalized by the optimal doping level ($p_c$, dashed vertical line) for each system. The calculations are done for a $256\times256$ $\mathbf{k}$-grid except for the mean-field case, for which a $512\times 512$ $\mathbf{k}$-grid is used.}
\end{figure}

Figure~\ref{fig4} presents the doping dependence of the ratio $2\Delta(0)/T_c$ for two different values of $U$.
When $U=0$, the ratio returns to the BCS value and remains nearly constant. 
For $U = 2$, in contrast, the ratio rises sharply as doping is reduced and becomes much larger than the BCS value in the underdoped regime. 
This corroborates previous numerical investigations of the intractable Hubbard model based on approximation schemes (red circles)~\cite{gull13superconductivity} and is further consistent with the experimentally observed doping dependence of this ratio in YBa$_2$Cu$_3$O$_{6+\delta}$ (YBCO) superconductors (red triangles)~\cite{inosov11crossover}.
In Fig.~\ref{fig4}, we also show the full mean-field ratio, where both the repulsive interaction $U$ and the superconducting pairing interaction are treated at the mean-field level. The gap is determined from the full mean-field gap equation, $(J/N)\sum_\mathbf{k}\int d\omega A(\mathbf{k}, \omega) g_\mathbf{k}^2 / (2\sqrt{\omega^2+\Delta^2(0)g_\mathbf{k}^2/4})=1$ and $T_c$ is calculated by Eq.~(\ref{eqTc}) with $A'=0$.
The full mean-field ratio for $U=2$ does not significantly differ from that for $U=0$ and is much smaller than that with the full consideration of the Fermi arcs in the underdoped regime.
This demonstrates that the enhancement of the gap-to-$T_c$ ratio is a consequence of the many-body nature (not the spectral property) of the Fermi arcs.


\section{Conclusion and discussion}
Using an exactly solvable Fermi-arc model, we showed analytically that exact incorporation of the Fermi arcs fundamentally modifies the BCS equations and isolated an additional many-body effect that suppresses $T_c$ beyond the simple reduction expected from a shrinking Fermi surface.
We further demonstrate that, in the pseudogap regime, this many-body effect leads to the formation of a new low-energy mode upon entering the superconducting phase, suppresses the superfluid stiffness, and substantially enhances the gap-to-$T_c$ ratio, all of which are consistent with experimental observations in cuprate superconductors.

Although the model considered in this work is sufficiently intricate to generate new insights into the interplay between Fermi arcs and superconductivity, and to reproduce several key features of cuprate superconductors, it does not exhibit dynamical spectral weight transfer~\cite{meinders93spectral,phillips10colloquium} in the absence of the pairing interaction. 
This limitation arises from the fact that the kinetic and potential energy operators in Eq.~(\ref{eqH}) commute.
Another limitation of the model is the explicit decoupling of the Coulomb repulsion from the pairing interaction, which implies that the microscopic origin of the pairing ``glue'' is not addressed within this framework.
It is possible to go beyond these limitations to gain a deeper understanding of high-$T_c$ superconductivity by employing more advanced exactly solvable models~\cite{mai26twisting}.

\begin{acknowledgments}
We thank Edwin Huang for helpful discussions.
This work was supported by the start-up grant from the Institute of Physics, Chinese Academy of Sciences. S.M. acknowledges financial support from Ministry of Science and Technology (Grant No. 2021YFA1400200), National Natural Science Fund of China (Grants No.12450401 and No. 12025407), and Chinese Academy of Sciences (No.YSBR047).
\end{acknowledgments}

\bibliography{refs.bib}

\end{document}